\documentclass[aps,prx,twocolumn,groupedaddress,amsmath,superscriptaddress]{revtex4-2}
\usepackage{amsfonts}
\usepackage{amsmath}
\usepackage[dvips]{graphicx}
\usepackage{color}
\usepackage{graphicx}
\usepackage{amssymb}
\usepackage{hyperref}
\usepackage{color}
\usepackage{mathrsfs}
\usepackage{amsthm}
\usepackage{epstopdf}
\usepackage{txfonts}
\usepackage{dsfont}
\usepackage{setspace}
\usepackage{multirow}
\usepackage{booktabs}
\usepackage{tabularx}
\usepackage{adjustbox}
\usepackage{colortbl}
\usepackage{hhline}
\usepackage{xcolor}
\usepackage{babel}
\usepackage{float}
\usepackage[normalem]{ulem}
\allowdisplaybreaks[4]

\makeatletter
\definecolor{newtxtcolor1}{rgb}{0.5,0,0.5}
\definecolor{newtxtcolor2}{rgb}{0.72,0.46,0.83}
\definecolor{newtxtcolor3}{rgb}{0.86,0.47,0.3}
\definecolor{nblue}{rgb}{0.3,0.3,1.0}%229
\definecolor{ngreen}{rgb}{0.2,0.7,0.2}%161
\definecolor{nred}{rgb}{0.9,0.1,0}%711&900
\definecolor{nblack}{rgb}{0,0,0}

\begin{document}
	
\title{Correlation-pattern-based Continuous-variable Entanglement Detection through Neural Networks}
\author{Xiaoting Gao}
\address{State Key Laboratory for Mesoscopic Physics, School of Physics, Frontiers Science Center for Nano-optoelectronics, Peking University, Beijing 100871, China}
\address{Laboratoire Kastler Brossel, Sorbonne Universit\'{e}, CNRS, ENS-PSL Research University, College de France, 4 place Jussieu, F-75252 Paris, France}
\author{Mathieu Isoard}
\address{Laboratoire Kastler Brossel, Sorbonne Universit\'{e}, CNRS, ENS-PSL Research University, College de France, 4 place Jussieu, F-75252 Paris, France}
\author{Fengxiao Sun}
\address{State Key Laboratory for Mesoscopic Physics, School of Physics, Frontiers Science Center for Nano-optoelectronics, Peking University, Beijing 100871, China}
\address{Collaborative Innovation Center of Extreme Optics, Shanxi University, Taiyuan, Shanxi 030006, China}
\author{Carlos E. Lopetegui}
\address{Laboratoire Kastler Brossel, Sorbonne Universit\'{e}, CNRS, ENS-PSL Research University, College de France, 4 place Jussieu, F-75252 Paris, France}
\author{Yu~Xiang}
\email{xiangy.phy@pku.edu.cn}
\address{State Key Laboratory for Mesoscopic Physics, School of Physics, Frontiers Science Center for Nano-optoelectronics, Peking University, Beijing 100871, China}
\address{Collaborative Innovation Center of Extreme Optics, Shanxi University, Taiyuan, Shanxi 030006, China}
\author{Valentina Parigi}
\address{Laboratoire Kastler Brossel, Sorbonne Universit\'{e}, CNRS, ENS-PSL Research University, College de France, 4 place Jussieu, F-75252 Paris, France}
\author{Qiongyi~He}
\email{qiongyihe@pku.edu.cn}
\address{State Key Laboratory for Mesoscopic Physics, School of Physics, Frontiers Science Center for Nano-optoelectronics, Peking University, Beijing 100871, China}
\address{Collaborative Innovation Center of Extreme Optics, Shanxi University, Taiyuan, Shanxi 030006, China}
\address{Peking University Yangtze Delta Institute of Optoelectronics, Nantong 226010, Jiangsu, China}		
\address{Hefei National Laboratory, Hefei 230088, China}
\author{Mattia Walschaers}
\email{mattia.walschaers@lkb.upmc.fr}
\address{Laboratoire Kastler Brossel, Sorbonne Universit\'{e}, CNRS, ENS-PSL Research University, College de France, 4 place Jussieu, F-75252 Paris, France}
	
\date{\today}
	
\begin{abstract}
Entanglement in continuous-variable non-Gaussian states provides irreplaceable advantages in many quantum information tasks. However, the sheer amount of information in such states grows exponentially and makes a full characterization impossible. Here, we develop a neural network that allows us to use correlation patterns to effectively detect continuous-variable entanglement through homodyne detection. Using a recently defined stellar hierarchy to rank the states used for training, our algorithm works not only on any kind of Gaussian state but also on a whole class of experimentally achievable non-Gaussian states, including photon-subtracted states. With the same limited amount of data, our method provides higher accuracy than usual methods to detect entanglement based on maximum-likelihood tomography. Moreover, in order to visualize the effect of the neural network, we employ a dimension reduction algorithm on the patterns. This shows that a clear boundary appears between the entangled states and others after the neural network processing. In addition, these techniques allow us to compare different entanglement witnesses and understand their working. Our findings provide a new approach for experimental detection of continuous-variable quantum correlations without resorting to a full tomography of the state and confirm the exciting potential of neural networks in quantum information processing.
\end{abstract}
\maketitle
%%%%%%%%%%%%%%%%%%%%%Introduction
	
\textit{Introduction.---}The study of quantum entanglement is experiencing a thorough theoretical development and an impressive experimental progress~\cite{1935,RevModPhys.81.865}, leading to important applications in quantum cryptography~\cite{PhysRevLett.67.661}, quantum metrology~\cite{Toth_2014} and quantum computation~\cite{Jozsa}. It is, therefore, crucial to find reliable and practical methods to detect entanglement. Especially in the continuous variable (CV) regime, significant breakthroughs have recently been achieved in the experimental preparation of non-Gaussian entanglement~\cite{Chang2020,ra2022non}. Such entangled states have been proven to be essential resource for entanglement distillation~\cite{eisert2002distilling,fiuravsek2002gaussian,nphoton.2010.1}, quantum-enhanced imaging~\cite{PhysRevApplied.16.064037,PhysRevResearch.4.043010} and sensing~\cite{science.345,PhysRevLett.107.083601}, and to reach a quantum computational advantage~\cite{PhysRevLett.130.090602}. However, entanglement detection in such complex systems turns out to be a challenging problem.
	
The conventional entanglement criteria which rely on the knowledge of reconstructed density matrix, such as the positive partial transpose (PPT) criterion~\cite{peres1996separability} or the quantum Fisher information (QFI) criterion proposed in Ref.~\cite{gessner2016efficient}, are experimentally infeasible for general non-Gaussian states. A common thought is to avoid the time-consuming process of two-mode tomography~\cite{lvovsky2009continuous}, which requires performing joint measurements on many quadrature combinations~\cite{PhysRevLett.127.150502}. Only for some states with specific analytic structures~\cite{ourjoumtsev2006quantum}, this demanding procedure can be simplified to a series of single-mode homodyne tomography~\cite{ourjoumtsev2007increasing}.
An innovative approach to overcome this issue is provided by deep neural networks~\cite{lecun2015deep}, which can work with limited amounts of data from actual measurements. Recently, neural networks have found extensive applications in quantum physics and quantum information science~\cite{RevModPhys.91.045002,gebhart2023learning}, including detecting quantum features~\cite{cimini2020neural,ahmed2021classification,urena2023entanglement,rieger_sample-efficient_2023}, characterizing quantum states and quantum channels~\cite{carrasquilla2019reconstructing,tiunov2020experimental,PhysRevApplied.18.044041,fedotova2022continuous,zhu2022flexible,PhysRevA.106.012409,zhu2023predictive}, and solving many-body problems~\cite{carleo2017solving,gao2017efficient,PRXQuantum.2.040201}. A key step thus lies in selecting an appropriate training data set to ensure that the networks can effectively and universally learn the features of the quantum system. Keeping our focus on the homodyne measurements which are feasible in CV experiments, we seek to answer the following question in this paper: Can neural networks be used to detect entanglement for general non-Gaussian states?

In this work, we develop a deep learning algorithm to detect entanglement for arbitrary two-mode CV states, including experimentally relevant photon subtracted states \cite{ra2022non}. Instead of extracting entanglement properties from the density matrices, our neural network is only based on four correlation patterns, which can be easily measured via homodyne measurements. It can be found that our algorithm achieves much higher accuracy than PPT and QFI criteria based on the maximum likelihood (MaxLik) algorithm with the same homodyne data. Our network also shows strong robustness for single-photon subtracted states. Furthermore, with a visualization tool, namely a t-SNE algorithm, we show that elusive entanglement information is hidden in the correlation patterns, hence in the joint probability distributions. It can be seen that the neural network is indeed able to correctly sort out data: clusters of entangled states clearly emerge after neural network processing. Therefore, our findings provide an approach for detecting CV entanglement in experimentally-accessible ways and confirm the deep neural network approach as a powerful tool in quantum information processing.
	
\begin{figure}[t] 
\includegraphics[width=8.5cm]{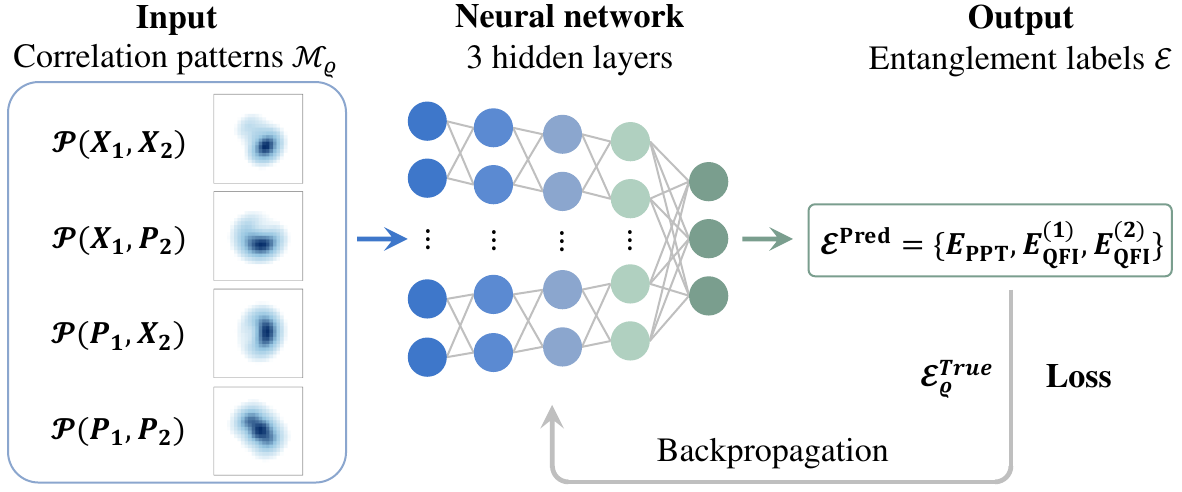}
\caption{\label{process} 
Scheme of the training data processing. The generation of the training data set begins with a series of random density matrices $\varrho$. Then for each density matrix one generates $24\times24\times4$-dimensional correlation patterns $\mathcal{M}_{\varrho}$ as input data of the neural network. At the output, $3$ entanglement labels $\mathcal{E}^\text{Pred}$ are computed from $\varrho$ and fed into the neural network for training. The loss function is evaluated between the true entanglement labels $\mathcal{E}_\varrho^{\text{True}}$ and the predicted labels $\mathcal{E}^{\text{Pred}}$ output from the neural network.
}
\end{figure}
	
\textit{Generation and selection of CV states.---}We start by generally characterizing two-mode CV quantum states in order to generate an abundant training data set. To do so, we rely on the recently developed stellar formalism~\cite{chabaud2020stellar,chabaud2021classical,chabaud2022holomorphic}. In this formalism, we analyse a pure state $|\psi\rangle$ in terms of its stellar function $F_\psi^{\star}(\mathbf{z})$. To define this function, we start by considering the state's decomposition in the Fock basis, i.e., $|\psi\rangle=\sum\limits_{\mathbf{n}\geq0}\psi_{\mathbf{n}}|\mathbf{n}\rangle\in\mathcal{H}^{\otimes 2}$ with $\mathbf{n}=(n_1,n_2)$, such that the stellar function can be written as
\begin{align}
F_\psi^{\star}(\mathbf{z}) \equiv e^{\frac{1}{2}\|\mathbf{z}\|^2}\left\langle \mathbf{z}^*|\psi\right\rangle=\sum\limits_{n_1,n_2} \frac{\psi_\mathbf{n}}{\sqrt{n_1!n_2!}} z_1^{n_1}z_2^{n_2},~~\forall~\mathbf{z}=(z_1,z_2)\in\mathbb{C}^2
\end{align}
where $|\mathbf{z}\rangle=\text{e}^{-\frac{1}{2}\|\mathbf{z}\|^2}\sum\limits_{n_1,n_2}\frac{z_1^{n_1}z_2^{n_2}}{\sqrt{n_1!n_2!}}|n_1,n_2\rangle$ is a coherent state of complex amplitude $\mathbf{z}$. The stellar rank $r$ of $|\psi\rangle$ is defined as the number of zeros of its stellar function, representing a minimal non-Gaussian operational cost. For instance, $r=0$ means that the state is Gaussian, while $r=1$ corresponds to a class of non-Gaussian states that contains, both, single-photon added and subtracted states~\cite{chabaud2022holomorphic}. Any multimode pure state $|\psi\rangle$ with finite stellar rank $r$ can be decomposed into $|\psi\rangle=\hat G|C\rangle$, where $\hat G$ is a given Gaussian operator acting onto the state $|C\rangle$, which is called core state; it is a normalized pure quantum state with multivariate polynomial stellar function of degree $r$, equal to the stellar rank of the state. It then follows immediately that Gaussian operations $\hat G$ must preserve the stellar rank~\cite{chabaud2021classical}. 
	
We generate states $\hat\rho$ in our data set by first creating a core state $|C\rangle$ with a given stellar rank $r\leq2$ and random complex coefficients for the superposition in the Fock basis.  Then, according to the Bloch-Messiah decomposition, any multimode Gaussian unitary operation $\hat G$ can be decomposed as $\hat G=\mathcal{\hat U}(\mathbf{\varphi}) \mathcal{\hat S}(\mathbf{\xi})\mathcal{\hat D}(\mathbf{\alpha})\mathcal{\hat V}(\mathbf{\phi})$, where $\mathcal{\hat U}$ and $\mathcal{\hat V}$ are beam-splitter operators, $\mathcal{\hat S}$ is a squeezing operator and $\mathcal{\hat D}$ is a displacement operator. We choose random values for the parameters $\mathbf{\varphi}$, $\mathbf{\xi}$, $\mathbf{\alpha}$ and $\mathbf{\phi}$ of the different operators and apply the corresponding operation $\hat G$ to the core state $|C\rangle$ to produce the random state $|\psi\rangle=\hat G|C\rangle$. Finally, by adding optical losses to simulate an experimental loss channel, a lossy two-mode state $\hat\rho$ is generated. More details can be found in \ref{App1}.

\textit{Homodyne data.---}The aim of our work is to feed our neural network with data that are directly accessible in experiments. To this goal, we focus on quadrature statistics, which in quantum optics experiments can be directly obtained through homodyne detection. The quadrature observables $\hat x_{k}$ and $\hat p_{k}$ in the modes $k= 1, 2$ can be defined as the real and imaginary parts of the photon annihilation operator $\hat a_k$ such that $\hat x_{k} \equiv (\hat a_k + \hat a_k^{\dag})$ and $\hat p_{k} \equiv i(\hat a_k^{\dag} - \hat a_k)$. Homodyne detection then corresponds to a projective measurement on eigenstates of these quadrature operators. Hence, we define these eigenstates as $\hat x_k|X_k\rangle = X_k |X_k\rangle$ and $\hat p_k|P_k\rangle = P_k |P_k\rangle$, where $X_k$ and $P_k$ describe the continuum of real measurement outcomes for the quadrature measurements in the mode $k$. 

For any given state $\hat \rho$, we can obtain the joint quadrature statistics $\mathcal{P}(X_1,X_2) \equiv \langle X_1; X_2 |\hat \rho |X_1; X_2\rangle$ (other joint statistics are defined analogously). The density matrix is known in its Fock basis decomposition, we can explicitly calculate the joint quadrature statistics as
\begin{equation}\label{eq:JointQuad}
\mathcal{P}(X_1,X_2) = \sum_{\substack{n_1, n_2 \\ n'_1, n'_2} \geq 0}^{N} \rho_{n_1,n_2;n'_1, n'_2} \langle X_1 | n_1 \rangle \langle X_2 | n_2 \rangle \langle X_1 | n'_1 \rangle^* \langle X_2 | n'_2 \rangle^*. 
\end{equation}
These quantities can be evaluated directly using the wave functions of Fock states which are given by $\langle X_k | n_k \rangle = H(n_k, X_k/\sqrt{2}) e^{-\frac{X_k^2}{4}}/[(2 \pi)^{1/4} \sqrt{2^{n_k} n_k!}]$, where $H(n_k, X_k/\sqrt{2})$ denotes a Hermite polynomial. Other joint quadrature distributions can be calculated analogously, using $\langle P_k | n_k \rangle = i^{n_k} H(n_k, P_k/\sqrt{2}) e^{-\frac{P_k^2}{4}}/[(2 \pi)^{1/4} \sqrt{2^{n_k} n_k!}]$.

\textit{Training process.---}The method described above to generate random states is repeated $15,000$ times to obtain a set $\varrho$ with a wide variety of density matrices. To evaluate the performance of the model on an independent data set, we use a cross-validation method to tune the hyperparameters of the model and split the entire data set into two parts: $70\%$ for training and $30\%$ for validation. For each density matrix $\hat\rho$, four joint quadrature statistics $\mathcal{P}(X_1,X_2)$, $\mathcal{P}(X_1,P_2)$, $\mathcal{P}(P_1,X_2)$, $\mathcal{P}(P_1,P_2)$, see Eq. \eqref{eq:JointQuad}, and the corresponding output entanglement label $\mathcal{E}_{\hat\rho}^{\text{True}}$ are calculated, see Fig,~\ref{process} for a scheme of the training data processing. Since the joint probability distributions are continuous over the whole phase space, they need to be discretized before we can feed the neural network with them. To that end, we restrict the region of phase space with values going from -6 to 6 and bin each distribution into a $24\times24$ correlation pattern. Every pixel is given by the median value of the joint probability distribution in the corresponding grid. For each state $\hat\rho$, we thus obtain a $24\times24\times4$-dimensional tensor $\mathcal{M}_{\hat\rho}$. Then for the full set of density matrices, $\mathcal{M}_{\varrho}$ together with the entanglement labels $\mathcal{E}^\text{True}_\varrho$ are used for training the neural network with Adam optimization algorithm.
As shown in Fig.~\ref{process}, each node of the three hidden fully connected layers performs a unique linear transformation on the input vector from its previous layer, followed by a nonlinear ReLU activation function. The loss function is evaluated between the true entanglement labels $\mathcal{E}_\varrho^{\text{True}}$ and the predicted labels $\mathcal{E}^{\text{Pred}}$ output from the neural network with the binary cross-entropy. 
%$L=-\left[\mathcal{E}_\varrho^{\text{True}}\text{log}\left(\mathcal{E}^{\text{Pred}}\right)+\left(1-\mathcal{E}_\varrho^{\text{True}}\right)\text{log}\left(1-\mathcal{E}^{\text{Pred}}\right)\right]$. 
The backward processes are iterated until the loss function converges.

The binary entanglement labels $\mathcal{E}^{\text{True}}_\varrho=\{E_{\text{PPT}}$, $E_{\text{QFI}}^{(1)}$, $E_{\text{QFI}}^{(2)}\}_\varrho$ are obtained for the classification task to detect whether the state in set $\varrho$ is entangled or not via PPT~\cite{peres1996separability} and QFI~\cite{gessner2016efficient} criteria. In PPT criterion, the two-mode state is entangled if the partial transpose over one of its modes has negative eigenvalues, which leads to $\lVert\hat{\rho}^{\rm{T}_B}\rVert_1 > 1$. Here, we label the states that satisfy this inequality as $E_{\rm{PPT}}=1$ and the rest as $E_{\rm{PPT}}=0$. The metrological-entanglement-based QFI criterion, is based on estimating a parameter $\theta$ which is implemented through $\hat \rho_{\theta} = e^{-i\theta \hat A}\hat \rho e^{i\theta \hat A}$, where $\hat{A}=\sum_{i=1}^{2}\hat{A}_i$ is a sum of arbitrary local observable $\hat{A}_i$ for the $i$th mode. The intuitive idea behind the QFI criterion is to detect entanglement by showing $\hat \rho$ allows us to estimate $\theta$ with a precision that is higher than what could have been achieved with a separable state with the same variances fr the generators. More rigorously, the criterion is given by $E[\hat\rho,\hat{A}]=F_Q(\hat\rho, \hat{A})-4\sum_{i=1}^{2}\text{Var}[\hat\rho,\hat{A}_i]$, where $F_Q$ is the quantum Fisher information of state $\hat\rho$, and $\text{Var}[ \cdot]$ denotes the variance. 
The entanglement witness depends on the choice of the observable $\hat{A}_i$, which can be constructed by optimizing over an arbitrary linear combination of operators (namely, $\hat{A}_i=\mathbf{n} \cdot \mathbf{H}=\sum_{k=1} n_k \hat{H}_k$)~\cite{gessner2016efficient}. At the first order, $\mathbf{H}$ takes the form of $\mathbf{H^{(1)}}=(\hat{x}_i,\hat{p}_i)$ with $i=1,2$. To capture more correlations of non-Gaussian states, we can extend the set of operators by adding three second-order nonlinear operators: $\mathbf{H^{(2)}}=(\hat x_i,\hat p_i,\hat x_i^2,\hat p_i^2,(\hat x_i\hat p_i+\hat p_i\hat x_i)/2)$. If $E[\hat\rho,\hat{A}]>0$, the state is identified as QFI-type entangled and labeled as $E_{\text{QFI}}^{(n)}=1$, otherwise it is labeled as $E_{\text{QFI}}^{(n)}=0$, where $n$ refers to the set $\mathbf{H}^{(n)}$ which is used to compute $E[\hat\rho,\hat A]$.

\begin{figure}[t]
\includegraphics[width=8.5cm]{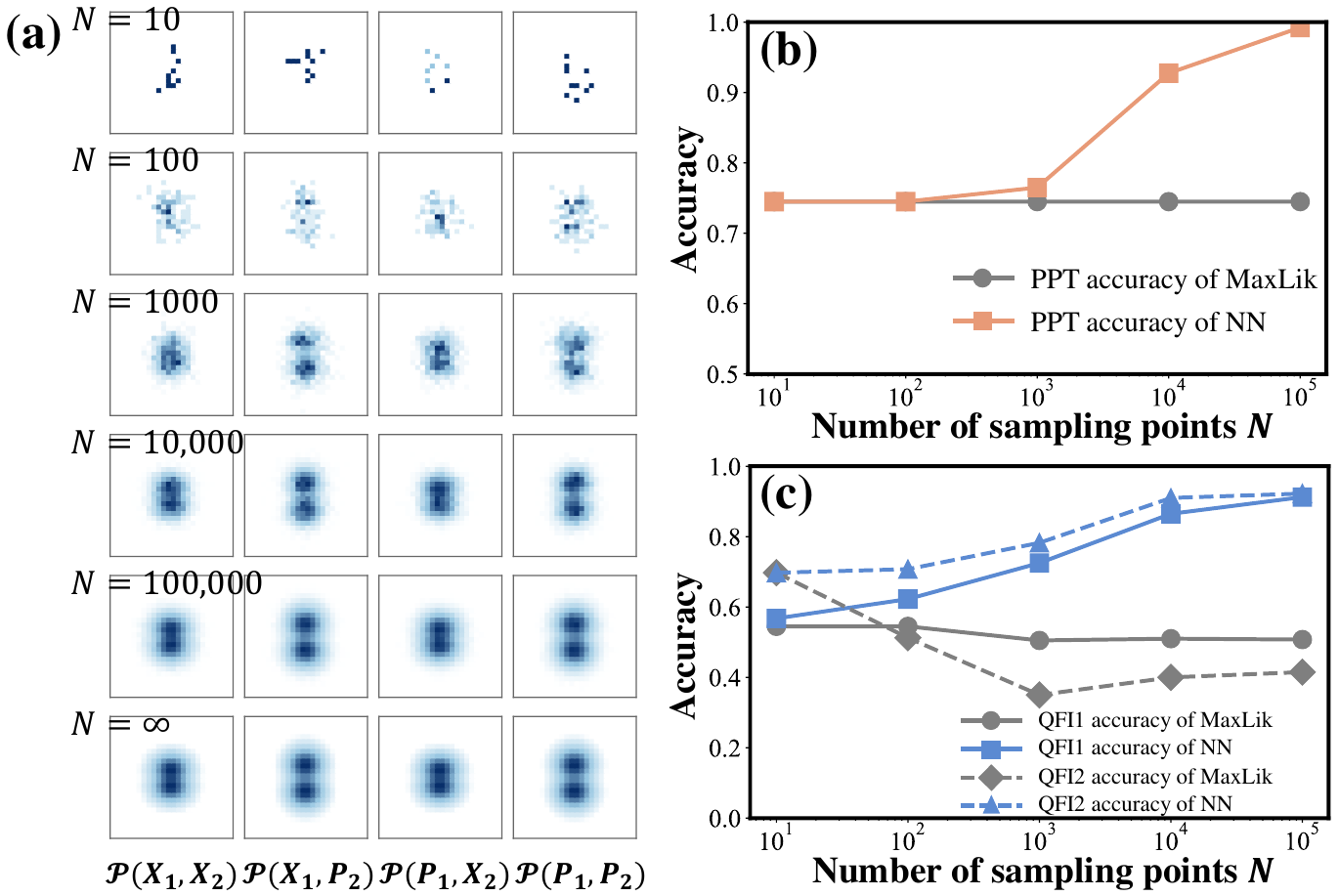}
\caption{\label{fig:epsart} 
(a)~Binned correlation patterns of a state $\hat\rho\in\varrho_{\text{test}}$ when the number of Monte Carlo sampling points $N=10,100,1000,10,000$ and $100,000$. The plot with $N=\infty$ shows the correlation patterns directly discretized from the theoretical joint probability distributions. (b) Accuracy of PPT-type entanglement prediction from the neural network (orange line) and MaxLik algorithm (gray line) against the same value of $N$ in (a). (c) Accuracy of QFI-type entanglement prediction from the neural network (blue lines) and MaxLik (gray lines) algorithm against $N$. Solid and dashed lines represent the first and second-order QFI, respectively.}
\label{acc}
\end{figure}
	
After $3,000$ epochs of training the loss function has converged and the network has captured the features mapping the correlation patterns to the entanglement labels, without providing the full density matrices $\varrho$. This is a crucial element since experiments generally do not have access to the full density matrix of a produced state, and what can be acquired is partial correlation information from measurements.
	
\textit{Testing process.---}To test the network with experimental-like data, we simulate the homodyne measurement outcomes via a Monte Carlo sampling method. The test data are obtained from previously unseen quantum states, denoted as $\varrho_{\text{test}}$. For each pattern of the state in $\varrho_{\text{test}}$, we perform $N$ repetitions of sampling to simulate the joint measurement events for each mode, forming a $2\times N$-dimensional outcomes and used to recover the joint probability distributions. However, directly feeding the raw sampling results into the neural network is infeasible, as the input layer of our trained network requires $24\times24\times4$-dimensional data. Thus, we also bin each $2\times N$ sampling points into a $24\times24$-dimensional matrix. Figure~\ref{acc}(a) shows the discretized correlation patterns with different numbers of sampling points $N$. The plot with $N=\infty$ is directly obtained from discretizing the theoretical joint probability distributions. As the number of samples $N$ increases from $10$ to $100,000$, the Monte Carlo sampling result converges towards the $N=\infty$ case.
	
\begin{figure*}[t]
\includegraphics[width=17cm]{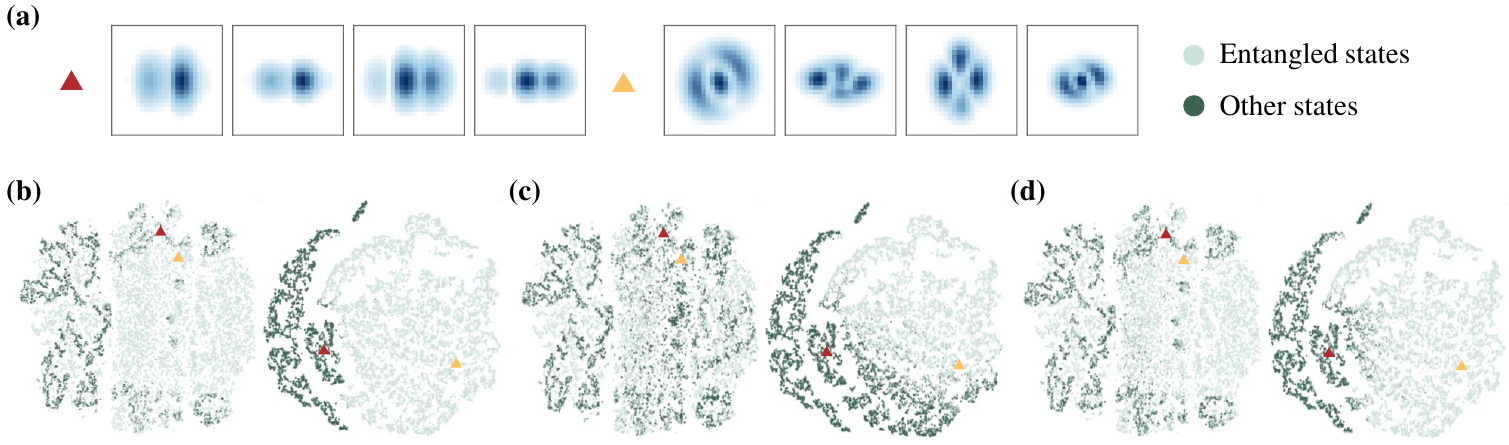}
\caption{\label{fig:epsart} 
Two-dimensional clusters of two-mode CV states. (a) Examples of $24\times24\times4$-dimensional correlation patterns for two input states. (b) Left: The 2-dimensional clustering of states before being fed into the network, where t-SNE preserves the pairwise similarities between data points. Right: The same dimension reduction process is conducted on the $64$-dimensional array from the last hidden layer of the neural network. Points representing PPT-type entangled ($E_{\text{PPT}}=1$) are colored in light green, while others are colored in deep green. (c) and (d) use the same method as (b) but are colored according to the first-order and second-order QFI-type entanglement labels $E_{\text{QFI}}^{(2)}$ and $E_{\text{QFI}}^{(2)}$, respectively.}
\label{cluster}
\end{figure*}
	
We compare the performance of the neural network with PPT and QFI entanglement predictions estimated from the MaxLik algorithm by quantifying the ratio of states the different algorithms correctly classified within the set $\varrho_{\text{test}}$. 
The MaxLik algorithm is a statistical method and can process the same Monte Carlo sampling outcomes to conduct the state tomography and reconstruct corresponding density matrices $\varrho_{\text{MaxLik}}$~\cite{lvovsky2009continuous}. After $20$ iterations of the algorithm, the entanglement labels $\mathcal{E}^{\text{MaxLik}}$ can be derived using PPT and QFI criteria. A significant difference between deep learning and MaxLik is that the former excels at extracting features and insights from a large data set, enabling the neural network to extrapolate and make accurate predictions for unseen data, while the latter reconstructs each state separately without relying on prior experience.
	
In Figs.~\ref{acc}(b) and (c), the orange and blue lines show the accuracy of $\mathcal{E}^{\text{Pred}}$ predicted by our neural network. The gray lines show the accuracy of $\mathcal{E}^{\text{MaxLik}}$ based on MaxLik. For PPT-type entanglement $E_{\text{PPT}}$, the accuracy of the neural network achieves 0.993 when the number of Monte Carlo sampling points is $N=100,000$. On the contrary, even though the average fidelity between $\varrho_{\text{test}}$ and the reconstructed density matrix $\varrho_{\text{MaxLik}}$ increases from 0.734 to 0.945, the accuracy of MaxLik remains unchanged when the number of samples increases. For QFI-type entanglement $E_{\text{QFI}}^{(n)}$, the accuracy of the neural network achieves 0.913 and 0.923 for the first and second order when $N=100,000$, respectively. With the same amount of joint homodyne measurement data, the neural network shows better performance than the standard state reconstruction process through the MaxLik algorithm.
	
%%%%%%%%%%%%%%%%%%%%%%Example: 1-sub state %%%%%%%%%%%%%%%%%%%%%%%%
Furthermore, we test our network on lossy single-photon subtracted states~\cite{ra2020non}. This class of states is included in our training data set  since it can be decomposed as
\begin{align}\label{1subexp}
|\psi\rangle_{\text{sub}}&\propto(\text{cos}\gamma \hat a_1+\text{sin}\gamma \hat a_2)\hat S_1(\xi_1)\hat S_2(\xi_2)|00\rangle\\\nonumber
&=\hat S_1(\xi_1)\hat S_2(\xi_2)(\text{cos}\gamma\text{sinh}r_1\text{e}^{i\omega_1}|10\rangle+\text{sin}\gamma\text{sinh}r_2\text{e}^{i\omega_2}|01\rangle).
\end{align}
where $\hat S_i(\xi_i)$ is the single-mode squeezing on mode $i$ with squeezing parameter $\xi_i=r_i\text{e}^{\omega_i}$, and $\gamma$ is the angle controlling the probability in which mode the single photon is subtracted. Hence we can also test how our network performs for this kind of state. The accuracy to predict PPT-type entanglement achieves $0.985$ when the number of sampling points is $N=100,000$. We compare the robustness of our network with another existing experiment-friendly criterion based on Fisher information reported in Ref.~\cite{barral2023metrological} for a state $|\psi\rangle_{\text{sub}}$ with squeezing parameters $r_1=2\text{dB}$ and $r_2=-3\text{dB}$, and $\omega_1=\omega_2=0$. While the other criterion cannot detect the entanglement when the loss coefficient $\eta$ is larger than 0.06, our neural network performs with a much stronger robustness leading to a loss threshold $\eta=0.33$ for the first-order QFI, and beyond $\eta=0.5$ for PPT and second order QFI (see Fig.~\ref{rho} of the Appendix). 
	
\textit{Clustering process.---}To visualise how our neural network processes input correlation patterns, we use t-Distributed Stochastic Neighbor Embedding (t-SNE)~\cite{van2008visualizing} as a dimension reduction technique to compare our data before and after the processing by the neural network.  Different from the supervised classifier neural networks, the t-SNE algorithm is an unsupervised learning approach dealing with unlabeled data. This visualization tool has been widely used in many biology and marketing problems, like cell population identification~\cite{kobak2019art,greener2022guide} and customer segmentation~\cite{Shen2021}. It can be seen that clusters of entangled states clearly emerge after neural network processing.

Figure~\ref{cluster}(a) shows two $24\times24\times4$-dimensional correlation patterns of a state with non-entangled labels $\{0,0,0\}$ and a state with entangled labels $\{1,1,1\}$, marked with red and yellow triangles, respectively. These high-dimensional data can be mapped into a two-dimensional plane through the t-SNE algorithm and the results are shown in Figs.~\ref{cluster}(b)(c)(d). The left plot of Fig.~\ref{cluster}(b) exhibits clusters formed from the discretized correlation patterns of the whole training data set ($n=15,000$). Each point represents a quantum state, colored by its PPT-type entanglement label $E_{\text{PPT}}=0$ (dark green) or $E_{\text{PPT}}=1$ (light green). The right plot of Fig.~\ref{cluster}(b) reveals clusters formed from output vectors of the last hidden layer of the neural network with $64$ neurons. After the neural network processing, the two overlapped dark green and light green clouds have largely detached from each other, forming disjoint clusters. We can clearly see that the two triangles significantly separate and now belong to their respective cluster.

Similarly, clusters in Figs.~\ref{cluster}(c) and (d) are of the same shape as in (b), but colored with the first-order and second-order QFI-type entanglement labels $E_{\text{QFI}}^{(1)}$ and $E_{\text{QFI}}^{(2)}$. Again, the two clouds are better separated after the neural network processing [right plots of (c) and (d)]. Comparing (c) with (d), we can see that the light green cluster in (d) covers more area than in (c), which intuitively shows that, as expected, second-order QFI-type criterion finds more entangled states than first-order. These results clearly show how different clusters of states have different metrological capabilities.
	
Even though there is no explicit boundary between the two classes (entangled or not) in the left cluster of Fig.~\ref{cluster}(b), where the neural network has not been applied yet, the two classes of states already tend to cluster together. This implies that the correlation patterns inherently contain entanglement-related regularities for the studied states, %thus the mapping from correlation patterns $\mathcal{M}_\varrho$ to entanglement labels $\mathcal{E}_\varrho$ is nontrivial, 
which makes it more feasible for the deep learning algorithm to learn from the training data. Therefore, this visualization method provides us with a technique to select appropriate input data when detecting a specific quantum feature through the deep learning algorithm.
	
%%%%%%%%%%%%%%%%%%%%%% Conclusion %%%%%%%%%%%%%%%%%%%%%%%%
In conclusion, we develop a neural network to detect CV entanglement for general two-mode states with finite stellar rank, only using correlation patterns obtained through homodyne detection. We test the performance of the network on patterns generated by Monte Carlo with different numbers of sampling points. With the same limited patterns, our method provides higher accuracy compared to the entanglement that can be extracted through a full state tomography. Meanwhile, the neural network shows strong robustness to losses, which we illustrate for the specific case of single-photon-subtracted states. Finally, we use the t-SNE algorithm to visualize the clusters formed from abundant correlation patterns before and after they are fed into the network. This helps us validate the suitability of the input for detecting target quantum features. This can be further used to identify and detect more refined kinds of quantum correlations, like not passive separability, i.e., the fact that the entanglement cannot be undone with optical passive transformations (beamsplitters, phase shifters), a strong feature of non-Gaussian states necessary to reach quantum advantage~\cite{Chabaud2023}.
	
\begin{acknowledgments}
We acknowledge enlightening discussions with Nicolas Treps. This work is supported by the National Natural Science Foundation of China (Grants No.~11975026, No. 12125402, No.~12004011, and No.~12147148), Beijing Natural Science Foundation (Grant No.~Z190005), and the Innovation Program for Quantum Science and Technology (Grant No.~2021ZD0301500). M.I., C.E.L., and M.W. received funding through the ANR JCJC project NoRdiC (ANR-21-CE47-0005). V.P. and M.W. acknowledge financial support from the European Research Council under the Consolidator Grant COQCOoN (Grant No. 820079).
\end{acknowledgments}

\bibliography{apssamp1017}

\appendix
\section{Generation of the data set}\label{App1}
\subsection{Generation of random quantum states}
\begin{figure*}[bht]
	\includegraphics[width=0.85\linewidth]{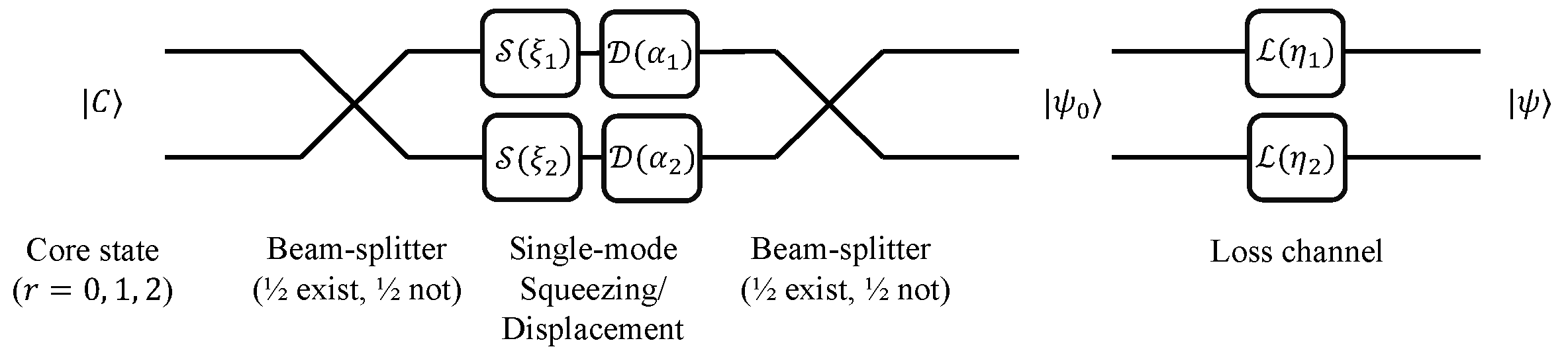}
	\caption{\label{circuit} The generation process of random states $\hat\rho$.}
	\label{rho}
\end{figure*}

The generation of a random state $\hat\rho$ in our data set begins with a core state $|C\rangle$ with a given stellar rank $r$ and random complex superposition coefficients of Fock basis. 
We restrict $r$ to $0, 1$ and $2$, which includes all Gaussian states and the most common two-mode non-Gaussian states in experiments like single-photon subtracted states. 
According to the Williamson decomposition and the Bloch-Messiah decomposition, a $2$-mode Gaussian unitary operation $\hat G$ can be decomposed as $\hat G=\mathcal{\hat U}(\varphi)\left( \prod\limits_{i=1}^2 \mathcal{\hat S}_i(\xi_i)\mathcal{\hat D}_i(\alpha_i)\right)\mathcal{\hat V}(\phi)$, where $\mathcal{\hat S}_i(\xi_i)=\text{e}^{\frac{1}{2}(\xi_i^*\hat a_i^2-\xi_i\hat a_i^{\dagger2})}$ is a squeezing operator with complex squeezing parameter $\xi_i$ acting on mode $i$ and $\mathcal{\hat D}_i(\alpha_i)=\text{e}^{\alpha_i\hat a^\dagger_i-\alpha_i^*\hat a_i}$ is a displacement operator with complex displacement amplitude $\alpha_i$ acting on mode $i$. 
Likewise, $\mathcal{\hat U}(\varphi)=\text{e}^{\varphi\hat a_1^\dagger \hat a_2-\varphi^*\hat a_1\hat a_2^\dagger}$ and $\mathcal{\hat V}(\phi)=\text{e}^{\phi\hat a_1^\dagger \hat a_2-\phi^*\hat a_1\hat a_2^\dagger}$ are beam-splitter operators, with complex coupling $\varphi$ and $\phi$. To obtain less entangled states to balance the proportion of the two classes, we set each of the two beam-splitter operators existing in the circuit with a $50\%$ probability. 
Losses are also added to the pure state
$|\psi\rangle=\hat G|C\rangle$
, using a single-mode loss channel $\hat L_i(\eta_i)$ with efficiency coefficient $\eta_i$ as described in Ref.~\cite{Eaton2022measurementbased}. 
The entire quantum circuit is shown in Fig.~\ref{circuit}, generating a two-mode state $\hat\rho$ with several randomly selected free parameters, given by
\begin{align}
\hat\rho=\left(\prod_{i=1}^2\hat L_i(\eta_i)\right)\hat G |C\rangle\langle C|\hat G^\dagger \left(\prod_{i=1}^2\hat L_i^\dagger(\eta_i)\right).
\end{align}\par
\par

\subsection{Binning of correlation patterns $\mathcal{M}_\varrho$}
\begin{figure*}[hbt]
	\includegraphics[width=13.6cm]{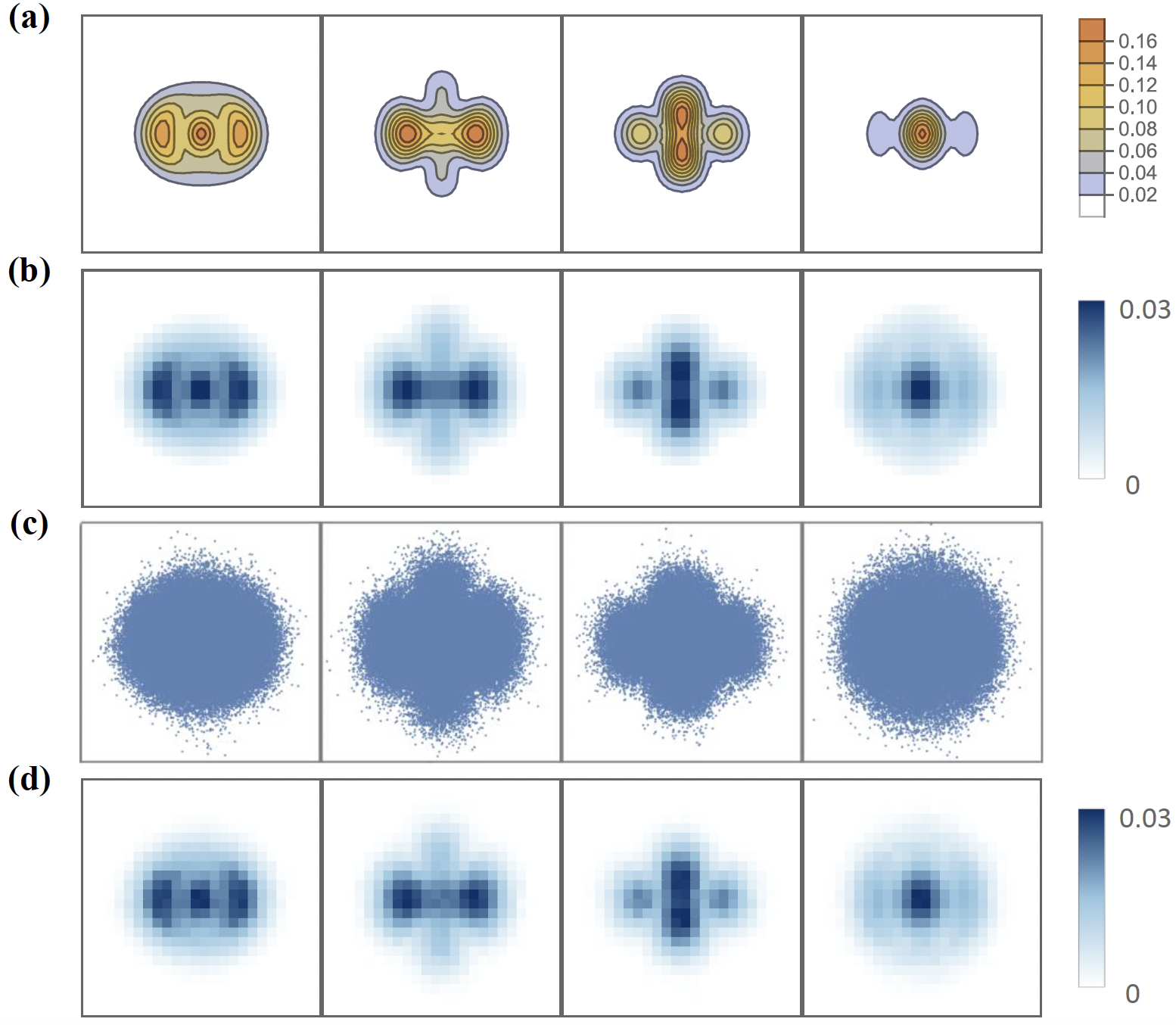}
	\caption{\label{3plots} The binning of input correlation patterns. (a)~Four theoretical joint probability distributions of a state. (b)~Binned correlation patterns of (a). (c)~Monte Carlo sampling results of the same state from its theoretical distributions. (d)~Binned correlation patterns of (c).}
	\label{rho}
\end{figure*}

The correlation pattern of the joint probability distribution $\mathcal{P}(X_1,X_2)$ extends in its continuous form over the whole phase space $X_1$ and $X_2$, as well as other distributions $\mathcal{P}(X_1,P_2)$, $\mathcal{P}(P_1,X_2)$ and $\mathcal{P}(P_1,P_2)$. 
Directly feeding them into neural networks is impossible, as the model requires limited and discretized data for effective processing. 
Therefore, we restrict the region of phase space from -6 to 6 and bin each distribution image into a normalized $24\times24$-dimensional matrix $m$. 
The matrix element $m_{jk}$ of distribution $\mathcal{P}(X_1,X_2)$ (other joint statistics are processed analogously) is given by its median value in the region $X_1\in(-6+\frac{1}{2}(k-1),-6+\frac{1}{2}k), X_2\in(6-\frac{1}{2}(j-1), 6-\frac{1}{2}j)$ to discretize the joint probability distribution, as shown in Fig.~\ref{3plots}. \par

Fig.~\ref{3plots}(a) shows four joint probability distributions of a state with entanglement label $\{E_{\text{PPT}}$, $E_{\text{QFI}}^{(1)}$, $E_{\text{QFI}}^{(2)}\}$=$\{1,0,1\}$, and their corresponding discretized correlation patterns can be seen in Fig.~\ref{3plots}(b). 
The Monte Carlo sampling results from theoretical joint probability distributions of the same state are shown in Fig.~\ref{3plots}(c). We sample $N=100,000$ points from each joint measurement distribution. The Monte Carlo algorithm simulates an actual homodyne sampling process. Binned correlation patterns of these samples are shown in Fig.~\ref{3plots}(d). \par
For network training, we use patterns of Fig.~\ref{3plots}(b) in order to save time for the sampling process. For testing, we use patterns binned from Monte Carlo as in Fig.~\ref{3plots}(d) to check how the neural network performs when feeding patterns from statistically independent sampling points. 

\subsection{Cut-off on entanglement labels $\mathcal{E}_\varrho$}
For the entanglement labels $\mathcal{E}^{\text{True}}_\varrho=\{E_{\text{PPT}}$, $E_{\text{QFI}}^{(1)}$, $E_{\text{QFI}}^{(2)}\}_\varrho$ of the training data, we set a cut-off value on each label to exclude limiting cases that are too close to zero and also to balance the proportion of the two classes. In this case, the PPT-type entanglement label for state $\hat\rho$ reads
\begin{align}
E^{\text{min}}_{\text{PPT}}[\hat\rho]=-\lambda_\text{min}(\hat\rho^{\text{T}_i}),
\end{align}
where $\lambda_\text{min}$ denotes the minimum eigenvalue and $\hat\rho^{\text{T}_i}$ is the partial transpose of $\hat\rho$ on mode $i$. 
If $E^{\text{min}}_{\text{PPT}}>10^{-3}$ holds, the state is labeled as $E_{\text{PPT}}=1$, otherwise it is labeled as $E_{\text{PPT}}=0$.\par 

The QFI-type entanglement witness given in the main text is
\begin{align}
E[\hat\rho,\hat{A}]=F_Q(\hat\rho,\hat{A})-4\left(\text{Var}[\hat\rho_1,\hat{A}_1]+\text{Var}[\hat\rho_2,\hat{A}_2]\right),
\end{align}
where $F_Q(\hat\rho,\hat{A})$ is the quantum Fisher information of state $\hat\rho$ with local generators $\hat{A}=\hat{A}_1+\hat{A}_2$ and $\text{Var}[\cdot]$ denotes the variance of an operator. 
If $E[\hat\rho,\hat{A}]>10^{-8}$ holds, the state is labeled as $E_{\text{QFI}}=1$, otherwise is labeled as $E_{\text{QFI}}=0$. The cut-off values are chosen to optimize to the performance of the neural network over the validation set.\par

%%%%%%%%%%%%%%%%%%%%%%%%%%%%%%%%%%%%%%%%%%%%%%
\begin{table*}[bht]
	\centering
	\caption{\label{tableNN}The Structure of the neural network.}
	\begin{tabular}{p{0.3\linewidth} p{0.2\linewidth}}
		\hline
		\textbf{Layers}& \textbf{Output shape}\\ 
		\hline
		Input layer& $24\times24\times4$\\
		Flatten layer& 2304\\
		Fully connected layer& 1024 \\
		ReLU activation layer& 1024\\
		Dropout& 1024\\
		Fully connected layer& 128\\
		ReLU activation layer& 128\\
		Dropout& 128\\
		Fully connected layer& 64\\
		ReLU activation layer& 64\\
		Dropout& 64\\
		Fully connected layer& 3\\
		Sigmoid activation layer& 3\\
		\hline
	\end{tabular}
\end{table*}
\section{Structure of the neural network}
As shown in Table.~\ref{tableNN}, the neural network consist of four fully connected layers with 1024, 128, 64 and 3 nodes, transforming the four discretized correlation patterns into three numbers. 
In a fully connected layer, all the neurons of the input are connected to every neuron of the output layer.
A weight matrix $W$ and bias vector $b$ are applied by each fully connected layer and perform a linear transformation $Wx+b$ on the input vector $x$ from the previous layer to the next one. 
Apart from the linear transformation, activation functions introduce non-linearity into the network. 
ReLU, which stands for Rectified Linear Unit, is one of the most widely used activation functions in artificial neural networks. 
It is mathematically defined as
\begin{align}
\lambda(y)=
\begin{cases}
y, & \text{if } y > 0 \\
0, & \text{if } y \leq 0
\end{cases}
\end{align}
when $y=Wx+b$ is inputted. 
The last sigmoid activation layer maps these numbers in the range of 0 to 1 through a nonlinear function $\sigma(y)=\frac{1}{1+\text{e}^{-y}}$. 
The output value is the probability of a sample belonging to one class for this binary classification problem. 
A threshold is set to classify the two kinds of states. 
Outputs $\geq0.5$ belong to the entangled class and $<0.5$ belong to the others.\par 

%%%%%%%%%%%%%%%%%%%%%%%%%%%%%%%%%%%%%%%%%%%%%%
\section{Robustness of the neural network}
\begin{figure}[hbt]
	\includegraphics[width=8.5cm]{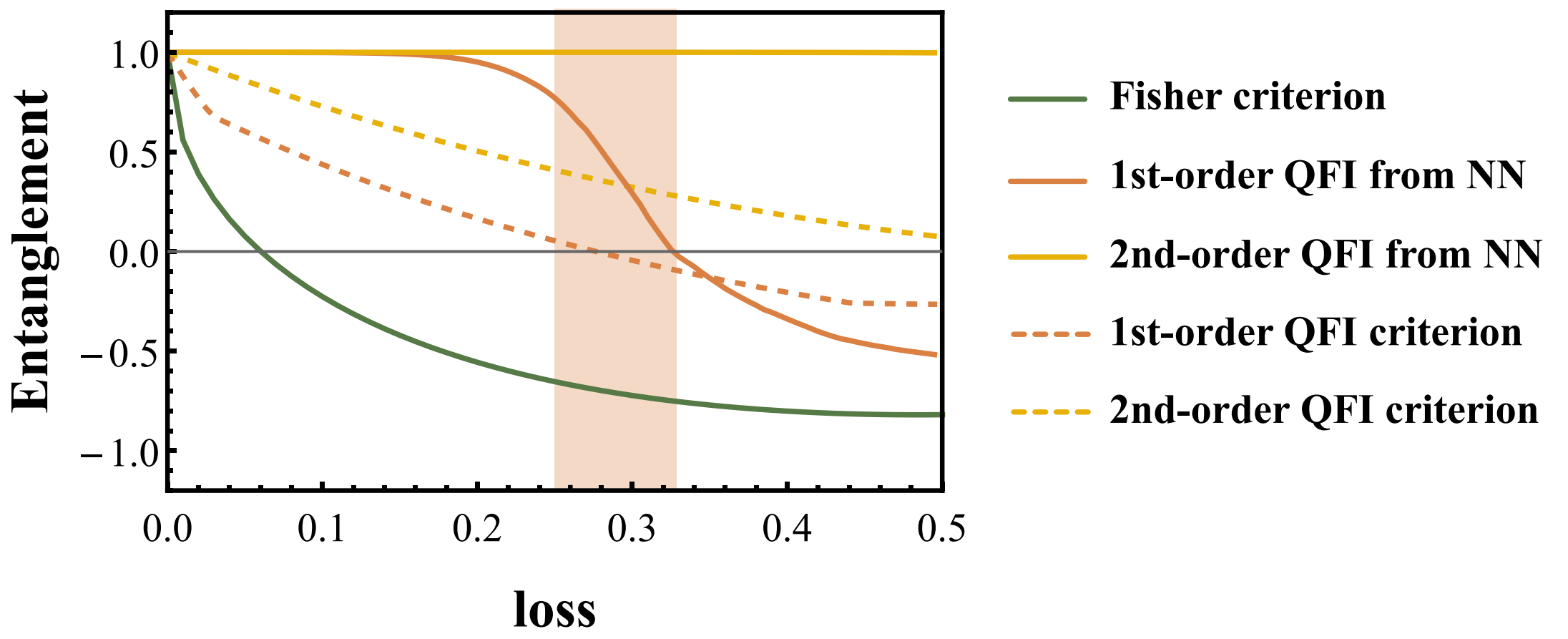}
	\caption{\label{1sub} Comparison of the robustness of Fisher information criterion, the neural network and quantum Fisher information criterion for a single-photon subtracted state with squeezing parameters $r_1=2\text{dB},r_2=-3\text{dB}$, squeezing phases $\omega_1=0,\omega_2=0$ and single-photon subtraction parameter $\gamma=\pi/4$ in Eq.~(3) in the main text. The positive value indicates entanglement can be witnessed by the corresponding criterion.}
	\label{rho}
\end{figure}

%\subsection{Single-photon subtracted states}
The expression of the single-photon subtracted state is given in Eq.~(3) in the main text. 
Fig.~\ref{1sub} shows the entanglement detection of a lossy single-photon subtracted state witnessed by the previous entanglement criterion based on Fisher information (FI) reported in Ref.~\cite{barral2023metrological}, compared with our neural network (NN) when providing the same homodyne measurement data. \par

The green line illustrates the entanglement witnessed by the FI criterion against losses. The orange and yellow solid lines represent the first-order and second-order QFI-type entanglement predictions at the output of the neural network. These outputs come from the final Sigmoid layer and fall within the range of 0 to 1, which represents the likelihood that a state is identified as entangled according to the respective criterion. In order to intuitively compare the entanglement detection results of the neural network with other criteria, we have normalized values from the network within the range of -1 to 1, where positive values indicate entanglement. 
Meanwhile, the orange and yellow dashed lines are the first-order and second-order QFI entanglement witness values calculated based on the state's density matrix.\par 
The thresholds of the predicted solid lines and theoretical dashed lines closely match. 
Indeed, the light red area corresponds to a region where the theoretical first-order QFI witness is very close to zero, meaning that the criterion hardly detects or does not detect entanglement. We observe that, similarly, this region is exactly where the output label of the neural network drops below 1 and reaches 0.5 (0 in Fig.~\ref{1sub}) at its right boundary; We recall that values of the label around 0.5 means that there is $50\%$ of chance that the state be entangled or not, and this tells us that the neural network does not witness clear features of entanglement for this state.
However, within the light red area in Fig.~\ref{1sub}, the neural network gives a probability of predicting QFI entanglement, while the results obtained directly from the density matrix (orange dashed line) go close to zero and even below.

\section{Principles of the t-SNE algorithm}
t-Distributed Stochastic Neighbor Embedding (t-SNE) is a kind of dimension reduction algorithm that maps the high-dimensional data to two or three-dimensional space~\cite{van2008visualizing}. 
The algorithm is divided into two parts. 
The first part is to create a probability distribution that represents similarities between neighbors for $N$ datapoints. 
The similarity of datapoint $x_j$ to datapoint $x_i$ is the conditional probability $p_{j|i}$ that $x_i$ would pick $x_j$ as its neighbor; it corresponds to the following Gaussian distribution:
\begin{align}
p_{j|i}=\frac{\text{exp}\left(-\|x_i-x_j\|^2/2\sigma_i^2\right)}{\sum\limits_{k\neq i}\left(-\|x_i-x_k\|^2/2\sigma_i^2\right)}.
\end{align}

The second part is to define a similar probability distribution in the low-dimensional space, using a Student t-distribution that has much heavier tails than the Gaussian distribution to solve the crowding problem:
\begin{align}
q_{ij} = \frac{\left(1 + \|y_i - y_j\|^2\right)^{-1}}{\sum\limits_{k \neq i}\left(1 + \|y_i - y_k\|^2\right)^{-1}}.
\end{align}
The location of point $y_i$ in the low-dimensional space is determined by minimizing the loss function given by Kullback-Leibler divergence of the distribution $p_{ij}=\frac{p_{j|i}+p_{i|j}}{2N}$ from the distribution $q_{ij}$
\begin{align}
KL(P||Q) = \sum_{i} p_{ij} \log\left(\frac{p_{ij}}{q_{ij}}\right).
\end{align}
This minimization process is performed using a gradient descent optimization algorithm. The optimization yields a map that reflects the similarities between the high-dimensional inputs.

% \clearpage
% \appendix
% \begin{widetext}
% \input{nn_Appendix.tex}
% \end{widetext}

\end{document}